\documentclass[ijoc,nonblindrev]{informs3}
\usepackage{color}
\usepackage{url}
\usepackage{hyperref}
 \usepackage{amsmath}
 \usepackage{multirow}
 \usepackage{tabularx}
\hypersetup{
    colorlinks=blue,
    linkcolor=blue,
    filecolor=blue,
    urlcolor=blue,
}
\OneAndAHalfSpacedXII 

\usepackage{natbib}
\usepackage[ruled]{algorithm}
\usepackage[noend]{algorithmic}
 \bibpunct[, ]{(}{)}{,}{a}{}{,}%
 \def\bibfont{\small}%
\TheoremsNumberedThrough     

\EquationsNumberedThrough    

\MANUSCRIPTNO{} 

\begin{document}

\TITLE{Submodular Optimization Beyond Nonnegativity: Adaptive Seed Selection in Incentivized Social Advertising}
\ARTICLEAUTHORS{%
\AUTHOR{Shaojie Tang}
\AFF{University of Texas at Dallas, \EMAIL{shaojie.tang@utdallas.edu}}
\AUTHOR{Jing Yuan}
\AFF{University of Texas at Dallas, \EMAIL{jing.yuan@utdallas.edu}}}

\ABSTRACT{The idea of social advertising (or social promotion) is to select a group of influential individuals (a.k.a \emph{seeds}) to help promote some products or ideas through an online social networks. There are two major players in the social advertising ecosystem: advertiser and platform. The platform sells viral engagements such as ``like''s to advertisers by inserting their ads into the feed of seeds. These seeds receive monetary incentives from the platform in exchange for their participation in the social advertising campaign. Once an ad is engaged by a follower of some seed, the platform receives a fixed amount of payment, called cost per engagement, from the advertiser. The ad could potentially attract more engagements from followers' followers and trigger a viral contagion. At the beginning of a campaign, the advertiser submits a budget to the platform and this budget can be used for two purposes: recruiting seeds and paying for the viral engagements generated by the seeds.  Note that the first part of payment goes to the seeds and the latter one is the actual revenue collected by the platform. In this setting, the problem for the platform is to recruit  a group of seeds such that she can collect the largest possible amount of  revenue subject to the budget constraint.  We formulate this problem as a seed selection problem whose objective function is defined as the minimum of the remaining budget after paying seeds and the cost of viral engagements generated by those seeds. This objective is non-monotone and it  might take on negative values, making existing results on submodular optimization and influence maximization not applicable to our setting. We study this problem under both non-adaptive and adaptive settings. Although we focus on social advertising in this paper, our results apply to any optimization problems whose objective function is the expectation of the  minimum of a stochastic submodular function and a linear function.}
\maketitle


\section{Introduction}
\label{sec-intro}
Social advertising (or social promotion) has been proved to be an effective approach that can produce a significant cascade of adoption through word-of-mouth effect.  It has been shown that social advertising is more effective than conventional advertising channels, including both demographically targeted and untargeted ads \cite{bakshy2012social,tucker2012social}. Social advertising is often implemented as \emph{promoted posts} that are displayed in the news feeds of their online users. Under the cost per engagement (CPE) pricing model, the advertiser pays the platform for all users engaged with their ad. Examples of engagement include ``like'', ``share'', or ``comment''. One unique feature of promoted posts, as compared with traditional online ads, is that they can be propagated from user to user. In particular, once a user $v$ engages with an ad, such an engagement will appear in the feed of $v$'s followers, who could be influenced to engage with the same ad. This potentially could trigger viral contagion.

In this paper, we study the revenue maximization problem in the context of incentivized social advertising. Our model involves two major players: advertiser and platform. At the beginning of a campaign, the advertiser submits a budget limit to the platform. The platform is in charge of running the campaign and planning budget on behalf of the advertiser. In particular, the platform  can spend the budget on behalf of the advertiser in two ways: recruiting seeds and paying for the viral engagements generated by the seeds. Note that the first part of the payment goes to the seeds and the latter one is the actual revenue collected by the platform. This motivates us to define the objective function of the platform as the minimum of the remaining budget after paying seeds and the value of viral engagements generated by those seeds. Formally, if the platform selects $S$ as seeds at cost $c(S)$ and it generates $g(S, \Phi)$ engagements at the end of the day, where $\Phi$ is a random variable capturing the uncertainty about the propagation of the engagements, we can represent the expected revenue $f_{exp}(S)$ of the platform as
\begin{eqnarray}
\label{eq:aaa1}
f_{exp}(S) = \mathbb{E}_{\Phi}[ \min \{\textsf{cost per engagement} \times g(S, \Phi), B-c(S) \}]
\end{eqnarray}

We study the revenue maximization seed selection problem from the platform's perspective under both non-adaptive and adaptive settings.  Under both settings, the platform's goal is to collect as many engagements as possible, while reducing the cost of hiring seed users.

\textbf{Our Results:} We next summarize the main contributions made in this paper.
\begin{itemize}
\item Under the non-adaptive setting, one must pick a group of seed users $S$ all at once in advance to  maximize $f_{exp}(S)$.   Notice that $f_{exp}(S)$ is non-monotone and it  might take on negative values, making most of existing results on submodular optimization and influence maximization not applicable to our setting. This problem becomes even more challenging by having to take into account the uncertainty about the spread of engagements through a social network. For the general non-adaptive seed selection problem, we develop an algorithm that achieves a $\frac{1-e^{-\frac{1}{2}}}{4}$ approximation ratio. For the case when the propagation of engagements is deterministic, we improve the approximation ratio to $\frac{1-e^{-1}}{2}$.

\item  Under the more complicated adaptive setting, one is allowed to pick seed users sequentially and adaptively, where each selection is based on the partial realizations of selected seeds (e.g., we choose the next seed given who we have selected as seeds so far, and how many engagements they have generated). Formally, we can encode any policy using a function from a set of
partial realizations to the set of users, specifying which seed to select next under a particular partial realization of selected seeds.  In this setting, our goal is to design a policy, rather than finding a fixed set of seeds, to maximize the expected revenue. We develop an adaptive strategy that achieves a $\alpha \frac{1-e^{-\frac{C}{B}}}{2}$ approximation ratio against the optimal adaptive policy where $B$ is the budget constraint and $C$, as well as $\alpha$, is some term depending on the cost of the most expensive seed. We show that if the cost of the most expensive seed is no larger than $B/2$, then the above approximation ratio is lower bounded by $\frac{1-e^{-\frac{1}{2}}}{4}$.
\item
Although we restrict our attention to incentivized social advertising in this paper, we make  fundamental contributions to the field of submodular optimization in that our results apply to any optimization problems whose objective function, which might take on negative values,  is the expectation of the  minimum of an increasing stochastic submodular function and a decreasing linear function.
\end{itemize}

\section{Related Work}
\paragraph{Influence Maximization and Social Advertising.} The problem of influence maximization has been studied in \cite{kempe2003maximizing} where their objective is to select a group of seed nodes to maximize the expected size of influence. They show that a simple greedy algorithm achieves a $(1-1/e-\epsilon)$ approximation ratio. \cite{golovin2011adaptive} extends this study to the adaptive setting where they develop an adaptive greedy policy that achieves a $(1-1/e-\epsilon)$ approximation ratio against the optimal adaptive policy. Unlike our model, their objective functions are monotone and (adaptive) submodular.   Our study is closely related to existing studies on social advertising. In \cite{chalermsook2015social}, they study the social advertising problem with multiple advertisers. However, they consider the cost of selecting seeds as sunk cost, i.e., each advertiser $i$ can select up to $k_i$ seeds where $k_i$ is pre-determined. Consequently, they concentrate on maximizing a monotone function subject to cardinality constraints. The channel allocation and user ordering problems are investigated in \cite{alon2012optimizing} and \cite{abbassi2015optimizing}, respectively. However, they do not consider engagement propagation.  In \cite{aslay2014viral}, they introduce the concept of regret to capture the tradeoff between maximizing the social advertising revenue and minimizing the impact of free-riding. Their goal is to select a group of seeds to minimize the regret. This problem is revisited in \cite{tang2016optimizing}, they convert it to a new optimization problem which admits constant approximation algorithms under some conditions. In \cite{aslay2016revenue}, they initiate the study of revenue maximization incentivized social advertising problem. They formulate their problem as a monotone submodular maximization problem subject to a partition matroid constraint and submodular knapsack constraints.  Although the basic business model adopted in their paper is similar to ours, we propose to use a new utility function to capture the revenue of the platform. Unlike the objective function defined in \cite{aslay2016revenue} which is monotone and submodular, our objective function is non-monotone and it  might take on negative values. This makes existing results on submodular optimization \cite{nemhauser1978analysis} and influence maximization \cite{kempe2003maximizing} not applicable to our setting.

\paragraph{Submodular Optimization.} Our study is also related to non-adaptive  \cite{nemhauser1978analysis} and adaptive submodular maximization \cite{golovin2011adaptive}. While most of existing studies in this field assume non-negative functions, \cite{harshaw2019submodular} \cite{sviridenko2017optimal} study the problem of maximizing a regularized submodular function, which may take on negative values, in the form of a sum of a non-negative monotone increasing submodular function and a linear function.  \cite{feldman2020guess} develop a faster algorithm using a surrogate objective that varies with time. For the case of a cardinality constraint and a non-positive linear part, \cite{harshaw2019submodular,kazemi2020regularized} develop the first practical algorithms. Our objective function is different from theirs, i.e., it is the expectation of the  minimum of an increasing stochastic submodular function and a decreasing linear function.  Moreover, we are the first to extend this study to the adaptive setting. We believe that our study makes fundamental contributions to the field of submodular optimization.

\section{Preliminaries and Problem Statement}
\subsection{Engagement Propagation Model and Submodularity}
\label{sec:1}
The platform owns a social network $G=(\mathcal{V}, \mathcal{E})$, where $\mathcal{V}$ represents a set of users and an edge $(u, v)\in \mathcal{E}$ means that user $v$ is a follower of user $u$, and
thus $v$ is exposed to $u$'s posts and may be influenced by $u$. We adopt the Independent Cascade Model (IC) \cite{kempe2003maximizing} to govern the way in which ad engagements (e.g., clicks and shares) propagate in $G$.  
Under the IC model, each edge $(u,v)\in \mathcal{E}$ has a influence probability $\rho_{uv}\in[0,1]$ which is the probability that $u$ influences $v$ to engage with the ad. The propagation process under IC can be described in discrete steps. Let $A_t$ denote the set of engaged users in step $t$. In step $t=0$, we select a group of seeds $A_0 = S$ to endorse the ad. Then, in each subsequence step $t$, each user $u \in A_t\setminus A_{t-1}$, that is newly engaged with the ad in step $t$, has a single chance to influence  each of its neighbors $v\in N(v) \cap (\mathcal{V}\setminus A_t)$ where $N(v)$ denotes the neighbors of $v$; it succeeds with a probability $\rho_{uv}$. If $u$ succeeds, then $v$ is added to $A_{t+1}$. Thus, for each user $v\in \mathcal{V}\setminus A_t$, $v$ becomes engaged with the ad with probability $1-\prod_{u\in N(v)\cap (A_t\setminus A_{t-1})} (1-\rho_{uv})$ at this step.  This process ends at a random step $T$ when no more users are newly engaged. The total number of engagements at the end of the campaign is $|\bigcup_{t\in[T]}A_t|$.

Alternatively, one can define the above engagement process over a live-edge graph $L(\mathcal{V}, L(\mathcal{E}))$. $L(\mathcal{V}, L(\mathcal{E}))$ is a random subgraph of $G$ and $L(\mathcal{E})$ is a subset of $\mathcal{E}$ which includes each edge $(u,v)\in \mathcal{E}$ independently with probability $\rho_{uv}$. All edges in $L(\mathcal{E})$ are labeled as ``live'' and the rest of them in $\mathcal{E}\setminus L(\mathcal{E})$ are labeled as ``blocked''. Given a set of seed users $S$ and a random graph  $L(\mathcal{V}, L(\mathcal{E}))$, let $M(S, L)$ denote the set of all users that can be reached by at least one user from $S$ through live edges. Informally, $M(S, L)$ is equivalent to the set of users which
are influenced by $S$. Thus, the total number of engagements of $S$ is $|M(S, L)|$. Given  a random graph  $L(\mathcal{V}, L(\mathcal{E}))$,  we model the state $\phi(u)$ of $u$ as a function $\phi(u): \mathcal{E}\rightarrow \{\verb"Blocked" , \verb"Live", ?\}$, where $\phi(u)((v,w))= \verb"Blocked"$ means that  $v$ can be reached by $u$ through live edges and the label of $(v,w)$ is blocked, $\phi(u)((v,w))=\verb"Live"$ means that  $v$ can be reached by $u$ through live edges and the label of $(v,w)$ is live, and $\phi(u)((v,w))=?$ means that  $v$ can not be reached by $u$ through live edges. Intuitively, $\phi(u)$ contains the labels of all edges whose labels can be observed after $u$ is being selected as a seed. Let $\phi=\{\phi(v)\mid v\in \mathcal{V}\}$ denote a \emph{full realization}. Notice that there is a one-to-one mapping between live-edge graphs and full realizations. Given a graph $G=(\mathcal{V}, \mathcal{E})$ and influence probability $\rho$,  let $\Phi$ denote a random realization, there is a known prior probability distribution $p(\phi)=\{\Pr[\Phi=\phi]: \phi\in \Omega\}$ over all possible realizations $\Omega$. For a given set of seeds $S$ and a realization $\phi$, we can represent the number of engagements $g(S, \phi)$ of $S$ conditional on $\phi$ as
\begin{eqnarray*}
  g(S, \phi)=|\{v\mid\exists u\in S, w\in \mathcal{V}\setminus S\mbox{ such that } \phi(u)((w, v))= \verb"Live" \}\cup S|
\end{eqnarray*}
Intuitively, $  g(S, \phi)$  is the number of users who can be reached by at least one user in $S$ through live edges (including $S$ itself) conditional on $\phi$. The expected number  of engagements $g_{exp}(S) $ of $S$ is
\begin{eqnarray*}
g_{exp}(S) =  \mathbb{E}[g(S, \Phi)]=\mathbb{E}[|\{v\mid \exists u\in S, w\in \mathcal{V}\setminus S\mbox{ such that } \Phi(u)((w, v))=\verb"Live" \}\cup S|]
\end{eqnarray*}
where the expectation is taken over $\Phi$ according to $p(\phi)$.  We next introduce the concept of submodularity and monotonicity.
\begin{definition}[Submodularity and Monotonicity] Consider two subsets $A\subseteq \mathcal{V}$ and $B \subseteq \mathcal{V}$ such that $A \subseteq B$ and a user $v\in \mathcal{V}\setminus B$, a function  $q: 2^{\mathcal{V}}\rightarrow \mathbb{R}_{\geq0}$ is submodular if  $q(A\cup\{v\})-q(A) \geq q(B\cup\{v\})-q(B)$.  A function  $q: 2^{\mathcal{V}}\rightarrow \mathbb{R}_{\geq0}$ is monotone if $q(A\cup\{v\})-q(A)\geq 0$.
\end{definition}

In \cite{kempe2003maximizing}, they show that $g(\cdot, \phi)$ is both monotone and submodular for any realization $\phi$ with $\Pr[\Phi=\phi]>0$.  Moreover, $g_{exp}(\cdot)$ is also monotone and submodular.

\subsection{Business Model}
\paragraph{\textbf{Advertiser.}} An advertiser submits a budget $B$ that specifies the maximum amount she is willing to pay on the campaign to the platform. This budget will be used for two purposes. For each seed user $v\in \mathcal{V}$ that is chosen by the platform to endorse the ad, the advertiser agrees to pay $v$ an incentive $c(v)$. She also agrees to pay each user that engages with  her ad a cost per engagement amount $cpe$. 

\paragraph{\textbf{Platform.}} The platform is hosting the campaign. After receiving a campaign budget $B$, as well as a cost-per-engagement amount $cpe$, from the advertiser, the platform selects a group of seed users $S\subseteq \mathcal{V}$ on behalf of the advertiser to endorse her ad, and each seed user $v\in S$ receives an incentive $c(v)$.
The total payment made by the advertiser is composed of two parts: The cost for the ad-engagements and the cost for hiring the seed users $c(S)=\sum_{v\in S} c(v)$. Notice that the revenue  of the platform is just the cost for the ad-engagements, as $c(S)$ is paid to the seed users. If there is no budget constraint, i.e, $B=\infty$, the revenue of the platform conditional on a realization $\phi$ is simply $cpe\cdot g(S, \phi)$ where $g(S, \phi)$ is the number of engagements generated by $S$ under $\phi$. 
For a general budget constraint $B$, we must ensure that the total cost of the ad-engagements is upper bounded by $B-c(S)$. Thus, we model the actual revenue $f(S, \phi)$ of the platform subject to a  budget constraint $B$ as $f(S, \phi) = \min \{cpe\cdot g(S, \phi), B-c(S) \}$. Without loss of generality, we normalize the value of  $cpe$ to one to obtain a simplified form of $f(S, \phi)$ as follows. 
\begin{eqnarray*}
f(S, \phi) = \min \{g(S, \phi), B-c(S) \}
\end{eqnarray*}
With the above notation, we can represent the expected revenue $f_{exp}(S)$ of $S$ as
\begin{eqnarray}
\label{eq:aaa}
f_{exp}(S) =\mathbb{E}[f(S, \Phi)] = \mathbb{E}[ \min \{g(S, \Phi), B-c(S) \}]
\end{eqnarray}
where the expectation is taken over $\Phi$ according to $p(\phi)$. In this paper, we study the revenue maximization seed selection problem under both non-adaptive setting and adaptive setting. Under the non-adaptive setting, one must pick a group of seed users all at once in advance. Under the more complicated adaptive setting, one is allowed to pick seed users sequentially and adaptively, where each selection is based on the feedback from the past observations. Under both settings, the platform faces the tradeoff of generating as many engagements as possible and reducing the cost of hiring influential seed users.
\subsection{Non-adaptive Seed Selection Problem}
We first describe the seed selection problem under the non-adaptive setting. Our objective is to  select a fixed set of seed users to maximize the expected revenue (\ref{eq:aaa}). Formally,
\[\max_{S\subseteq \mathcal{V}}\{f_{exp}(S)\mid c(S) \leq B\}\]
 One can drop the constraint from the above formulation without affecting its optimal solution. This is because if $S$ is an optimal solution to the above problem, then it must satisfy $f_{exp}(S)\geq 0$, which implies $c(S) \leq B$, otherwise, we can pick $\emptyset$ as a better solution which contradicts to the assumption that $S$ is an optimal solution. Notice that the utility function $f_{exp}(\cdot)$ might take on negative values, which renders the existing results on submodular optimization and influence maximization ineffective.

\subsection{Adaptive Seed Selection Problem}
Under the adaptive setting, the solution becomes a policy rather than a fixed set. We follow the framework of \cite{golovin2011adaptive} to introduce some notations. Formally, we can represent any policy using a mapping function $\pi$ that  maps a set of partial realizations  to $\mathcal{V}$: $\pi: 2^{\mathcal{V}}\times O^\mathcal{V} \rightarrow \mathcal{V}$ where $O = \{\verb"Blocked", \verb"Live", ?\}^\mathcal{E}$ denotes the set of all possible states of a user. Intuitively, a policy specifies which seed user to select next after observing a partial realization. Consider the following example for an illustration. Assume the current observation is $\{u, \phi(u)=\{(u,v) \mbox{ is }\verb"Live"\}, \mbox{ all other edges are } ?\}$, i.e., $u$ is the only selected seed and she has influenced $v$. If $\pi(\{u, \phi(u)=\{(u,v) \mbox{ is }\verb"Live", \mbox{ all other edges are } ?\}\})=w$, then $\pi$ selects $w$ as the next seed user. One is also allowed to design a randomized policy that maps a partial realization to a distribution of users. Since every randomized policy can be represented as a distribution of a set of deterministic policies, thus we focus on deterministic policies in this paper.

Let $\mathcal{V}(\pi, \phi)$ denote the set of seed users selected by $\pi$ conditional on a  realization $\phi$. Then the expected  utility $f_{avg}(\pi)$ of a policy $\pi$ can be written as
\begin{eqnarray}
f_{avg}(\pi)=\mathbb{E}[f(\mathcal{V}(\pi, \Phi), \Phi)]~\nonumber
\end{eqnarray}
where the expectation is taken over $\Phi$ according to $p(\phi)$. With the above notations,
the adaptive seed selection problem can be formulated as follows:

\[\max_{\pi}\{f_{avg}(\pi)\mid c(\mathcal{V}(\pi, \phi))\leq B \mbox{ for all realizations } \phi \mbox{ such that }\Pr[\Phi=\phi]>0\}\]

Again, one can drop the constraint from the above formulation without affecting its optimal solution. We next introduce some additional notations from \cite{golovin2011adaptive}.  Given any $S\subseteq \mathcal{V}$, let $\psi =\{\phi(v)\mid v\in S\}$ denote a \emph{partial realization} (e.g., $\psi$ encodes who we have selected as seeds and who have been influenced by them) and $\mathrm{dom}(\psi)=S$ is the \emph{domain} of $\psi$. Given a partial  realization $\psi$ and a realization $\phi$, we say $\phi$ is consistent with $\psi$, denoted $\phi \sim \psi$, if they are consistent everywhere in $\mathrm{dom}(\psi)$. A partial realization $\psi$  is said to be a \emph{subrealization} of  $\psi'$, denoted  $\psi \subseteq \psi'$, if $\mathrm{dom}(\psi) \subseteq \mathrm{dom}(\psi')$ and they are consistent everywhere in the domain $\mathrm{dom}(\psi)$ of $\psi$. Now we are ready to introduce the concept of adaptive submodularity and adaptive monotonicity.  

\begin{definition}\cite{golovin2011adaptive}[Adaptive Submodularity and Adaptive Monotonicity] Consider any two partial realizations $\psi$ and $\psi'$ such that $\psi\subseteq \psi'$. A function   $q: 2^{\mathcal{V}}\times O^\mathcal{V}\rightarrow \mathbb{R}_{\geq0}$  is called adaptive submodular if for each  $v\in \mathcal{V}\setminus \mathrm{dom}(\psi')$, we have
\[\mathbb{E}[q(\mathrm{dom}(\psi)\cup \{v\}, \Phi)-q(\mathrm{dom}(\psi), \Phi)] \geq \mathbb{E}[q(\mathrm{dom}(\psi')\cup \{v\}, \Phi)-q(\mathrm{dom}(\psi'), \Phi)]
\]
where the first expectation is taken over $\Phi$ with respect to $p(\phi\mid \psi)=\Pr(\Phi=\phi \mid \Phi \sim \psi)$ and the second expectation is taken over $\Phi$ with respect to $p(\phi\mid \psi')=\Pr(\Phi=\phi \mid \Phi \sim \psi')$.
 A function   $q: 2^{\mathcal{V}}\times O^\mathcal{V}\rightarrow \mathbb{R}_{\geq0}$ is called adaptive monotone if for all $\psi$ and $v\in \mathcal{V}\setminus \mathrm{dom}(\psi)$, we have $\mathbb{E}[q(\mathrm{dom}(\psi)\cup \{v\}, \Phi)-q(\mathrm{dom}(\psi), \Phi)] \geq 0$ where the expectation is taken over $\Phi$ with respect to $p(\phi\mid \psi)=\Pr(\Phi=\phi \mid \Phi \sim \psi)$.
\end{definition}

\section{Non-Adaptive Seed Selection Problem}
We first study a relaxed version of the non-adaptive seed selection  problem by assuming that the cost $c(S^*)$ of the optimal solution $S^*$ is known. Although the value of $c(S^*)$ is rarely known in practise, we start with this simplified case to make it easier to explain the idea of our approach for solving the general problem without knowing  $c(S^*)$. We also develop enhanced results for the case when there is only one realization, i.e., this happens when the influence probability of each edge is either $0$ or $1$.
\subsection{Warming Up: Seed Selection with Known $c(S^*)$}
To facilitate our algorithm design, we first introduce a new utility function. For any $z\in[0, B]$, define  $l(\cdot, z): 2^{\mathcal{V}}\rightarrow \mathbb{R}_{\geq0}$ as follows:
\begin{eqnarray*}
l(S, z) = \mathbb{E}[\min\{g(S, \Phi), B-z\}]
\end{eqnarray*}

We first show that  $l(\cdot, z): 2^{\mathcal{V}}\rightarrow \mathbb{R}_{\geq0}$ is monotone and submodular for any $z\in[0, B]$.
\begin{lemma}
\label{lem:000}
For any $z\in[0, B]$, $l(\cdot, c(S^*)):2^\mathcal{V}\rightarrow \mathbb{R}_{\geq0}$ is monotone and submodular.
\end{lemma}
\emph{Proof:}  It has been proved in \cite{kempe2003maximizing} that $g(\cdot, \phi)$ is monotone and submodular for any fixed realization $\phi$, thus $\min\{g(S, \phi), B-z\}$ is also monotone and submodular for any $z\in [0, B]$. This is because monotone submodular functions remain so under \emph{truncation} \cite{krause2014submodular} , i.e.,  the minimum of any monotone and submodular function and any constant is still monotone and submodular. Moreover, because submodularity is preserved under taking nonnegative linear combinations, we have $\mathbb{E}[\min\{g(S, \Phi), B-z\}]$ is also monotone and submodular for any $z\in [0, B]$. This finishes the proof of this lemma. $\Box$

In the rest of this paper, for any $x\in \mathbb{R}_{\geq0}$, let $\mathcal{V}(x)=\{e\in \mathcal{V}\mid c(e)\leq x\}$ denote the set of users whose cost is no larger than $x$. Assume $c(S^*)$ is given, we  introduce a new optimization problem \textbf{P.1} as follows.
\begin{center}
\framebox[0.78\textwidth][c]{
\enspace
\begin{minipage}[t]{0.45\textwidth}
\small
\textbf{P.1:} \emph{Maximize $l(S, c(S^*))$}\\
\textbf{subject to:}
\begin{equation*}
\begin{cases}
S\subseteq \mathcal{V}(c(S^*))\\
c(S) \leq c(S^*)
\end{cases}
\end{equation*}
\end{minipage}
}
\end{center}
Because $l(\cdot, c(S^*)): 2^\mathcal{V}\rightarrow \mathbb{R}_{\geq0}$ is submodular and monotone (Lemma \ref{lem:000}), \textbf{P.1} is a classical monotone submodular maximization problem subject to a knapsack constraint. To solve this problem efficiently, we consider two candidate solutions. One is the  singleton $e^*$ maximizing $l(\cdot, c(S^*)): 2^\mathcal{V}\rightarrow \mathbb{R}_{\geq0}$ among all users in $\mathcal{V}(c(S^*))$, i.e., $e^* = \argmax_{e\in \mathcal{V}(c(S^*))} l(\{e\}, c(S^*))$, and the other one is the output from a benefit-cost greedy algorithm $\textsf{Greedy}(c(S^*), c(S^*))$ listed in Algorithm \ref{alg:LPP2}. Note that  \textsf{Greedy}$(x, y)$ is presented as a general template that takes two parameters, i.e., $x$ and $y$, and this template makes it easier to describe our solution for solving the general problem later. Intuitively, \textsf{Greedy}$(x, y)$ refers to a benefit-cost greedy algorithm that runs on a ground set $\mathcal{V}(x)$ using the utility function $l(\cdot, y)$.  We next describe $\textsf{Greedy}(c(S^*), c(S^*))$  in details. It starts with iteration $t=0$ and the initial solution $S_0=\emptyset$, and in each subsequent iteration $t+1$, it adds $s_{t+1}$ to the current solution $S_t$, i.e., $S_{t+1}\leftarrow  S_{t} \cup\{s_{t+1}\}
$, where
\begin{eqnarray*}
s_{t+1}= \argmax_{e\in \mathcal{V}(c(S^*))\setminus S_t} \frac{l(S_t\cup\{e\}, c(S^*))-l(S_t, c(S^*))}{c(e)}
 \end{eqnarray*} denotes the user that maximizes the benefit-cost ratio with respect to $S_t$. This process iterates until it reaches some iteration $t$ such that $c(S_{t+1})+c(s_{t+1})> c(S^*)$.
\cite{khuller1999budgeted} show that  the better one between the above two candidates achieves an  approximation ratio of $(1-1/e)/2$ for maximizing a monotone and submodular function.

\begin{algorithm}[hptb]
\caption{\textsf{Greedy}$(x, y)$}
\label{alg:LPP2}
\begin{algorithmic}[1]
\STATE $S_0=\emptyset, t=0, U=\mathcal{V}(x)$
\WHILE {$U\setminus S_{t}\neq \emptyset$}
\STATE let $s_{t+1}$ denote the user $e$ maximizing $\frac{l(S_t\cup\{e\},  y)-l(S_t,  y)}{c(e)}$ among all users in $U\setminus S_{t}$
\IF {$ c(S_{t+1})+c(s_{t+1})\leq x$}
\STATE let $S_{t+1}=S_{t}\cup\{s_{t+1}\}$
\STATE $t\leftarrow t+1$
\ELSE
\RETURN $S_t$
\ENDIF
\ENDWHILE
\RETURN $S_{t}$
\end{algorithmic}
\end{algorithm}

\begin{lemma}\cite{khuller1999budgeted}
\label{lem:aaa}
Let $S^{\textsf{Greedy}(c(S^*), c(S^*))}$ denote the solution returned from \textsf{Greedy}($c(S^*), c(S^*)$) and let $S^{p1}$ denote the optimal solution to  \textbf{P.1},  $\max\{l(S^{\textsf{Greedy}(c(S^*), c(S^*))},  c(S^*)), l(\{e^*\}, c(S^*))\} \geq \frac{1-1/e}{2} l(S^{p1},c(S^*))$.
\end{lemma}

Next, we present the main theorem of this section where we show that the better one between $S^{\textsf{Greedy}(c(S^*),c(S^*))}$ and $\{e^*\}$ achieves a $\frac{1-1/e}{2}$ approximation ratio for our original problem.
\begin{theorem}
\label{thm:1}
Let $S^{\textsf{Greedy}(c(S^*), c(S^*))}$ denote the solution returned from \textsf{Greedy}($c(S^*), c(S^*)$), $\max\{f_{exp}(S^{\textsf{Greedy}(c(S^*), c(S^*))}), f_{exp}(\{e^*\})\} \geq \frac{1-1/e}{2} f_{exp}(S^*)$.
\end{theorem}
\emph{Proof:} Let $R=\argmax_{S\in\{S^{\textsf{Greedy}(c(S^*), c(S^*))}, \{e^*\}\} }l(S,  c(S^*))$ denote the solution maximizing $l(\cdot,  c(S^*))$ between $S^{\textsf{Greedy}(c(S^*), c(S^*))}$ and  $\{e^*\}$.  Since $f_{exp}(R)\leq \max\{f_{exp}(S^{\textsf{Greedy}(c(S^*), c(S^*))}), f_{exp}(\{e^*\})\}$, to prove this theorem, it suffice to show that $f_{exp}(R) \geq  \frac{1-1/e}{2} f_{exp}(S^*)$.

First, Lemma \ref{lem:aaa} implies that
\begin{eqnarray}
\label{eq:bbb}
l(R, c(S^*)) \geq \frac{1-1/e}{2} l(S^{p1},c(S^*)) \geq \frac{1-1/e}{2} l(S^*,c(S^*)) = \frac{1-1/e}{2} f_{exp}(S^*)
 \end{eqnarray}
The second inequality is due to $S^*$ is a feasible solution to \textbf{P.1} and $S^{p1}$ is the optimal solution to \textbf{P.1}. The equality is due to  the definition of $l(\cdot, c(S^*)):2^\mathcal{V}\rightarrow \mathbb{R}_{\geq0}$.

 Second, due to $c(e^*) \leq c(S^*)$, which is because $e^*\in S^*$, and $c(S^{\textsf{Greedy}(c(S^*), c(S^*))})\leq c(S^*)$, which is because of the design of Algorithm \ref{alg:LPP2}, we have $B-c(e^*) \geq B-c(S^*)$ and $B-c(S^{\textsf{Greedy}(c(S^*), c(S^*))}) \geq B-c(S^*)$. Thus,
 \begin{eqnarray}
\label{eq:ccc}B-c(R) \geq B-c(S^*)
 \end{eqnarray}
 Then we have $f_{exp}(R)=\mathbb{E}[\min\{g(R, \Phi), B-c(R)\}] \geq \mathbb{E}[\min\{g(R, \Phi), B-c(S^*)\}] = l(R,c(S^*)) \geq \frac{1-1/e}{2} f_{exp}(S^*)$ where the first inequality is due to (\ref{eq:ccc}) and the second inequality is due to (\ref{eq:bbb}). $\Box$

\begin{algorithm}[hptb]
\caption{Non-adaptive Seed Selection Algorithm}
\label{alg:LPP3}
\begin{algorithmic}[1]
\STATE run \textsf{Greedy}($\frac{B}{2}, 0$) and obtain an output $S^{\textsf{Greedy}(\frac{B}{2}, 0)}$
\STATE let $S^{phase1}= \argmax_{S\in \{S^{\textsf{Greedy}(\frac{B}{2}, 0)}, v(\frac{B}{2}, 0)\}} f_{exp}(S)$
\STATE let $\mathcal{V}_{large}= \mathcal{V}\setminus \mathcal{V}(\frac{B}{2})$
\FOR { $e\in \mathcal{V}_{large}$}
\STATE  run \textsf{Greedy}($(c(e), c(e))$) and obtain an output $S^{\textsf{Greedy}(c(e), c(e))}$
\ENDFOR
\STATE  let  $S^{phase2}$ be the best solution in $\bigcup_{ e\in \mathcal{V}_{large}}\{S^{\textsf{Greedy}(c(e), c(e))}, v(c(e), c(e))\}$
\RETURN the better solution between  $S^{phase1}$ and $S^{phase2}$
\end{algorithmic}
\end{algorithm}
\subsection{Solving the General Seed Selection Problem}
\label{sec:unconstrained}
Now we are in position to drop the assumption about knowing the value of $c(S^*)$. One naive approach of dealing with unknown $c(S^*)$ is to try all possibilities of $c(S^*)$. Then we feed each ``guess'' to Algorithm \ref{alg:LPP2} to obtain an output. At last, we choose the output maximizing the expected utility among all returned solutions as the final solution. This approach can secure an approximation ratio no worse than $\frac{1-1/e}{2}$. However, the number of possible values of  $c(S^*)$ is exponentially large in terms of $n$, it is clearly not affordable to enumerate all those possibilities. Perhaps surprisingly, we next show that it is not necessary to find out the value of $c(S^*)$, rather, we only need to find out the cost of the ``most expensive'' user in the optimal solution $S^*$, i.e., $\max_{e\in S^*} c(e)$. This can be done in $O(n)$ time since there are only $n$ possibilities of this value. Our algorithm (Algorithm \ref{alg:LPP3}) is composed of two phases. The first phase is dealing with the case when $\max_{e\in S^*} c(e)\leq \frac{B}{2}$, i.e., the cost of the most expensive seed from the optimal solution is no larger than $B/2$, and the second phase is used to handle the rest of the cases. At last, we return the solution maximizing the expected utility among all candidate outputs.

Before describing the design of our algorithm in details, we introduce some notations. For any $x, y \in \mathbb{R}_{\geq0}$, let $v(x, y)\in \argmax_{e\in \mathcal{V}(x)} l(\{e\}, y)$ denote the singleton maximizing $l(\cdot, y)$ among users in $\mathcal{V}(x)$. Recall that for any $x\in \mathbb{R}_{\geq0}$, we use $\mathcal{V}(x)=\{e\in \mathcal{V}\mid c(e)\leq x\}$ to denote the set of users whose cost is no larger than $x$, and  \textsf{Greedy}$(x, y)$ (Algorithm \ref{alg:LPP2}) refers to a benefit-cost greedy algorithm that runs on a ground set $\mathcal{V}(x)$ using the utility function $l(\cdot, y)$. Now we are in position to describe our algorithm.

\paragraph{Phase 1} 
Run \textsf{Greedy}($\frac{B}{2}, 0$) to obtain $S^{\textsf{Greedy}(\frac{B}{2}, 0)}$. Let $S^{phase1}$ be the better solution in $\{S^{\textsf{Greedy}(\frac{B}{2}, 0)}, v(\frac{B}{2},0)\}$.

\paragraph{Phase 2} Let $\mathcal{V}_{large}= \mathcal{V}\setminus \mathcal{V}(\frac{B}{2})$ denote the set of all users whose cost is larger than $\frac{B}{2}$.  For each $e\in \mathcal{V}_{large}$, run \textsf{Greedy}($(c(e), c(e))$) to obtain $S^{\textsf{Greedy}(c(e), c(e))}$. Let $S^{phase2}$ be the best solution in $\bigcup_{ e\in \mathcal{V}_{large}}\{S^{\textsf{Greedy}(c(e), c(e))}, v(c(e), c(e))\}$.

\paragraph{Output} Return the better solution between $S^{phase1}$ and $S^{phase2}$ as the final output.

We next analyze the approximation ratio of our algorithm. We first present two technical lemmas.  In the first lemma, we show that if $\max_{e\in S^*} c(e)> \frac{B}{2}$, then the solution returned from the second phase of our algorithm is near optimal. In the second  lemma, we show that if $\max_{e\in S^*} c(e)\leq \frac{B}{2}$, then the solution returned from the first phase of our algorithm is near optimal. Combining these two lemmas, we are able to derive an approximation bound of our algorithm for the original problem.

\begin{lemma}
\label{lem:rainy1}
If $\max_{e\in S^*} c(e)> \frac{B}{2}$, $f_{exp}(S^{phase2}) \geq  \frac{1-e^{-\frac{1}{2}}}{2} f_{exp}(S^*)$.
\end{lemma}
\emph{Proof:} Let $e' \in \argmax_{e\in S^*} c(e)$ denote the most expensive user in the optimal solution $S^*$. To prove this lemma, we first introduce a new optimization problem \textbf{P.3}.
\begin{center}
\framebox[0.78\textwidth][c]{
\enspace
\begin{minipage}[t]{0.45\textwidth}
\small
\textbf{P.3:} \emph{Maximize $l(S, c(e'))$}\\
\textbf{subject to:}
\begin{equation*}
\begin{cases}
S\subseteq \mathcal{V}(c(e'))\\
c(S) \leq B
\end{cases}
\end{equation*}
\end{minipage}
}
\end{center}
Because $\max_{e\in S^*} c(e)\leq c(e')$, we have $S^*\subseteq \mathcal{V}(c(e'))$. Thus, $S^*$ is a feasible solution to \textbf{P.3}. Moreover, we have  $l(S^*, c(e')) = \mathbb{E}[\min\{g(S^*, \Phi), B-c(e')\}]\geq \mathbb{E}[\min\{g(S^*, \Phi), B-c(S^*)\}]=f_{exp}(S^*)$ where the  inequality is due to $e'\in S^*$.  Let $S^{p3}$ denote the optimal solution to \textbf{P.3}, then we have $l(S^{p3}, c(e'))\geq l(S^*, c(e'))\geq f_{exp}(S^*)$. To prove this lemma, it suffice to show that $f_{exp}(S^{phase2}) \geq \frac{1-e^{-\frac{1}{2}}}{2} l(S^{p3}, c(e'))$.

Note that because $l(\cdot, c(e'))$ is monotone and submodular  (due to the same proof of Lemma \ref{lem:000}), \textbf{P.3} is a classical monotone submodular maximization problem subject to a knapsack constraint. 
Let $R=\argmax_{S\in \{S^{\textsf{Greedy}(c(e'), c(e'))}, \{v(c(e'), c(e'))\}\}}l(S, c(e'))$ denote the solution maximizing $l(\cdot, c(e'))$ between $S^{\textsf{Greedy}(c(e'), c(e'))}$ and $\{v(c(e'), c(e'))\}$, we next show that $l(R,c(e'))\geq \frac{1-e^{-\frac{c(e')}{B}}}{2}l(S^{p2}, c(e'))$. Consider a ``one-step-further'' version $\textsf{Greedy}^+(c(e'), c(e'))$ of $\textsf{Greedy}(c(e'), c(e'))$ which is obtained by first running $\textsf{Greedy}(c(e'), c(e'))$ then selecting one more user according to the same greedy manner. One can verify that $\textsf{Greedy}^+(c(e'), c(e'))$ always violates the budget constraint $c(e')$ (assuming $\mathcal{V}(c(e'))$ contains more than one users to avoid trivial cases). Let $S^{\textsf{Greedy}^+(c(e'), c(e'))}$ denote the solution returned from $\textsf{Greedy}^+(c(e'), c(e'))$. According to Theorem 2\footnote{The original theorem provides a stronger result than what we need here. I.e., they show that this result holds even under the adaptive setting.} in \cite{tang2020influence}, if $l(\cdot,c(e'))$ is monotone and submodular, then
\begin{eqnarray}
\label{eq:lastlast11}
 l(S^{\textsf{Greedy}^+(c(e'), c(e'))},c(e'))\geq 1-e^{-\frac{c(e')}{B}}l(S^{p2}, c(e'))
 \end{eqnarray}
Because $l(\cdot,c(e'))$ is monotone and submodular, the marginal utility brought by the last added user in $S^{\textsf{Greedy}^+(c(e'), c(e'))}$ is no larger than expected utility  of the best singleton $v(c(e'), c(e'))$. Thus,
\begin{eqnarray}
\label{eq:cold}
l(S^{\textsf{Greedy}^+(c(e'), c(e'))},c(e')) \leq l(S^{\textsf{Greedy}(c(e'), c(e'))}, c(e')) + l(\{v(c(e'), c(e'))\}, c(e'))
 \end{eqnarray}

 This implies that
\begin{eqnarray}
l(R,c(e'))&&\geq \frac{l(S^{\textsf{Greedy}(c(e'), c(e'))}, c(e')) + l(\{v(c(e'), c(e'))\}, c(e'))}{2} ~\nonumber\\
&&\geq \frac{l(S^{\textsf{Greedy}^+(c(e'), c(e'))},c(e'))}{2} \geq \frac{1-e^{-\frac{c(e')}{B}}}{2}l(S^{p2}, c(e'))
\label{eq:hott}
 \end{eqnarray}
The first inequality is due to the definition of $R$. The second inequality is due to (\ref{eq:cold}). The third inequality is due to (\ref{eq:lastlast11}).
Then we have
\begin{eqnarray}
l(R,c(e'))\geq \frac{1-e^{-\frac{c(e')}{B}}}{2}l(S^{p2}, c(e')) \geq \frac{1-e^{-\frac{1}{2}}}{2}l(S^{p2}, c(e'))
\label{eq:hot}
 \end{eqnarray}
The second inequality is due to $c(e')> \frac{B}{2}$.  Second, due to $c(v(c(e')))\leq c(e')$, which is because $v(c(e'), c(e')) \in \mathcal{V}(c(e'))$, and $c(S^{\textsf{Greedy}(c(e'), c(e'))})\leq c(e')$, which is because of the design of $\textsf{Greedy}(c(e'), c(e'))$, we have $B-c(v(c(e'), c(e'))) \geq B-c(S^*)$ and $B-c(S^{\textsf{Greedy}(c(e'), c(e'))}) \geq B-c(S^*)$. Thus,
 \begin{eqnarray}
 \label{eq:bread}
B-c(R) \geq B-c(S^*)
 \end{eqnarray}
(\ref{eq:hot}) and (\ref{eq:bread}) imply that $f_{exp}(R)=\mathbb{E}[\min\{g(R, \Phi), B-c(R)\}] \geq \mathbb{E}[\min\{g(R, \Phi),  B-c(S^*)\}]    = l(R,c(e')) \geq \frac{1-e^{-\frac{1}{2}}}{2}l(S^{p3}, c(e'))$. Note that $S^{\textsf{Greedy}(c(e'), c(e'))}$ and $\{v(c(e'), c(e'))\}$ are two of the
solutions considered in Phase 2  for its output $S^{phase2}$. Thus, we have $f_{exp}(S^{phase2}) \geq f_{exp}(R) \geq \frac{1-e^{-\frac{1}{2}}}{2} l(S^{p3}, c(e'))$. $\Box$

\begin{lemma}
\label{lem:rainy}
If $\max_{e\in S^*} c(e)\leq \frac{B}{2}$, $f_{exp}(S^{phase1}) \geq \frac{1-e^{-\frac{1}{2}}}{4} f_{exp}(S^*)$.
\end{lemma}
\emph{Proof:} To prove this lemma, we first introduce a new optimization problem \textbf{P.2}.
\begin{center}
\framebox[0.78\textwidth][c]{
\enspace
\begin{minipage}[t]{0.45\textwidth}
\small
\textbf{P.2:} \emph{Maximize $l(S, 0)$}\\
\textbf{subject to:}
\begin{equation*}
\begin{cases}
S\subseteq \mathcal{V}(\frac{B}{2})\\
c(S) \leq B
\end{cases}
\end{equation*}
\end{minipage}
}
\end{center}
If $\max_{e\in S^*} c(e)\leq \frac{B}{2}$, i.e., the cost of every seed in $S^*$ is no larger than $B/2$,  then we have $S^* \in \mathcal{V}(\frac{B}{2})$. It follows that $S^*$ is a feasible solution to \textbf{P.2}. Moreover, we have  $l(S^*, 0) = \mathbb{E}[\min\{g(S^*, \Phi), B\}]\geq \mathbb{E}[\min\{g(S^*, \Phi), B-c(S^*)\}]=f_{exp}(S^*)$.  Let $S^{p2}$ denote the optimal solution to \textbf{P.2}, then we have $l(S^{p2}, 0)\geq l(S^*, 0)\geq f_{exp}(S^*)$ where the first inequality is due to $S^{p2}$ is the optimal solution to \textbf{P.2} and  $S^*$ is a feasible solution to \textbf{P.2}. To prove this lemma, it suffice to show that $f_{exp}(S^{phase1}) \geq \frac{1-e^{-\frac{1}{2}}}{4} l(S^{p2}, 0)$ which is equivalent to showing that $\max\{f_{exp}(S^{\textsf{Greedy}(\frac{B}{2},0)}), f_{exp}(v(\frac{B}{2},0))\} \geq \frac{1-e^{-\frac{1}{2}}}{4} l(S^{p2}, 0)$.

Let $R=\argmax_{S\in \{S^{\textsf{Greedy}(\frac{B}{2},0)}, \{v(\frac{B}{2}, 0)\}\}}l(S, 0)$ denote the solution maximizing $l(\cdot, 0)$ between $S^{\textsf{Greedy}(\frac{B}{2}, 0)}$ and $\{v(\frac{B}{2},0)\}$. Note that because $l(\cdot, 0)$ is monotone and submodular (due to the same proof of Lemma \ref{lem:000}), \textbf{P.2} is a classical monotone submodular maximization problem subject to a knapsack constraint. With the same proof of (\ref{eq:hot}), we can show that
\begin{eqnarray}
\label{eq:uiwei}
l(R,0)\geq \frac{1-e^{-\frac{1}{2}}}{2}l(S^{p2}, 0)
\end{eqnarray}
Second, due to $c(v(\frac{B}{2},0)) \leq \frac{B}{2}$, which is because $v(\frac{B}{2},0) \in \mathcal{V}(\frac{B}{2})$, and $c(S^{\textsf{Greedy}(\frac{B}{2},0)})\leq \frac{B}{2}$, which is because of the design of \textsf{Greedy}($\frac{B}{2}, 0$), we have $B-c(v(\frac{B}{2},0)) \geq B-\frac{B}{2}$ and $B-c(S^{\textsf{Greedy}(\frac{B}{2},0)}) \geq B-\frac{B}{2}$. Thus,
 \begin{eqnarray}
 \label{eq:uiwei1}
B-c(R) \geq B-\frac{B}{2}=\frac{B}{2}
 \end{eqnarray}
 Then we have $f_{exp}(R)=\mathbb{E}[\min\{g(R, \Phi), B-c(R)\}] \geq \mathbb{E}[\min\{g(R, \Phi), \frac{B}{2}\}]  \geq \frac{1}{2}\mathbb{E}[\min\{g(R, \Phi), B\}]  = \frac{1}{2}l(R,0) \geq \frac{1-e^{-\frac{1}{2}}}{4}l(S^{p2}, 0)$ where the first inequality is due to (\ref{eq:uiwei1}) and the second inequality is due to (\ref{eq:uiwei}). $\Box$

Lemma \ref{lem:rainy} and Lemma \ref{lem:rainy1} together imply that $\max\{f_{exp}(S^{phase1}), f_{exp}(S^{phase2})\} \geq \frac{1-e^{-\frac{1}{2}}}{4} f_{exp}(S^*)$. Thus, we have the following main theorem.

\begin{theorem}
\label{thm:121}
The non-adaptive seed selection algorithm (Algorithm \ref{alg:LPP3}) achieves a $\frac{1-e^{-\frac{1}{2}}}{4}$ approximation ratio.
\end{theorem}

\subsection{Enhanced Results for Deterministic Realization}
Now we consider a special case of our problem when the realization is deterministic, i.e., there is only one realization $\phi$ with $\Pr[\Phi = \phi]=1$. Note that this happens if the propagation probability of each edge is either $0$ or $1$. For this special setting, we develop a new algorithm which achieves a $\frac{1-e^{-1}}{2}$ approximation ratio.  As there is only one realization, we have that for any $S\subseteq \mathcal{V}$, $f_{exp}(S)=f(S, \phi)$ and $g_{exp}(S) = g(S, \phi)$. Then the non-adaptive seed selection problem can be formulated as
\[ \max_{S: c(S)\leq B} \min\{g_{exp}(S), B-c(S)\} \]
Following the same flow of analysis used in the previous section, we first analyze the case when $c(S^*)$ is known, then drop this assumption in a later section.
\paragraph{Warming Up: Seed Selection with Known $c(S^*)$}
We first study a relaxed version of the non-adaptive social advertising problem by assuming that the cost $c(S^*)$ of the optimal solution $S^*$ is known. Recall that we use $\mathcal{V}(c(S^*))=\{e\in \mathcal{V}\mid c(e)\leq c(S^*)\}$ to denote the set of users whose cost is no larger than $c(S^*)$. We first introduce a new optimization problem \textbf{P.4} as follows.

\begin{center}
\framebox[0.78\textwidth][c]{
\enspace
\begin{minipage}[t]{0.45\textwidth}
\small
\textbf{P.4:} \emph{Maximize $g_{exp}(S)$}\\
\textbf{subject to:}
\begin{equation*}
\begin{cases}
S\subseteq \mathcal{V}(c(S^*))\\
c(S) \leq c(S^*)
\end{cases}
\end{equation*}
\end{minipage}
}
\end{center}
Because $g_{exp}(\cdot)$ is submodular and monotone, \textbf{P.4} is a classical monotone submodular maximization problem subject to a knapsack constraint. We still consider two candidate solutions. One is the best  singleton $e^*$ among all users in $\mathcal{V}(c(S^*))$, i.e., $e^* = \argmax_{e\in \mathcal{V}(c(S^*))} f_{exp}(\{e\})$, and the other one is the output from a benefit-cost greedy algorithm (Algorithm \ref{alg:LPP12}): It starts with iteration $t=0$ and initial solution $S_0=\emptyset$, and in each subsequent  iteration $t+1$, it adds $s_{t+1}$ to the existing solution $S_t$, i.e., $S_{t+1}\leftarrow  S_{t} \cup\{s_{t+1}\}$, where
\begin{eqnarray*}
s_{t+1}= \argmax_{e\in \mathcal{V}(c(S^*))\setminus S_t} \frac{g_{exp}(S_t\cup\{e\})-g_{exp}(S_t)}{c(e)}
\end{eqnarray*}
where denotes the user that maximizes the benefit-cost ratio with respect to $S_t$. This process iterates until it reaches some iteration $t$ such that $c(S_{t+1})+c(s_{t+1})> c(S^*)$.

Again, we first present a lemma from \cite{khuller1999budgeted} which states that the solution maximizing $g_{exp}(\cdot)$ between $\{e^*\}$ and $S^{alg3}$ achieves a $\frac{1-1/e}{2}$ approximation ratio for \textbf{P.4}.

\begin{algorithm}[hptb]
\caption{Benefit-cost Greedy Algorithm}
\label{alg:LPP12}
\begin{algorithmic}[1]
\STATE $S_0=\emptyset, t=0, U = \mathcal{V}(c(S^*))$
\WHILE {$U\setminus S_{t} \neq \emptyset$}
\STATE let $s_{t+1}$ denote the user $e$ maximizing $\frac{g_{exp}(S_t\cup\{e\})-g_{exp}(S_t)}{c(e)}$ among all users in $U\setminus S_{t}$
\IF {$ c(S_{t+1})+c(s_{t+1})\leq c(S^*)$}
\STATE let $S_{t+1}=S_{t}\cup\{s_{t+1}\}$
\STATE $t\leftarrow t+1$
\ELSE
\RETURN $S_t$
\ENDIF
\ENDWHILE
\RETURN $S_t$
\end{algorithmic}
\end{algorithm}

\begin{lemma}\cite{khuller1999budgeted}
\label{lem:aaa1}
Let $S^{alg3}$ denote the solution returned from Algorithm \ref{alg:LPP12} and let $S^{p4}$ denote the optimal solution to \textbf{P.4},  $\max\{g_{exp}(S^{alg3}), g_{exp}(\{e^*\})\} \geq \frac{1-1/e}{2} g_{exp}(S^{p4})$.
\end{lemma}

We next show that the solution maximizing $f_{exp}(\cdot)$ between $\{e^*\}$ and $S^{alg3}$ achieves a $\frac{1-1/e}{2}$ approximation ratio for our original problem.
\begin{theorem}
\label{thm:11}
$\max\{f_{exp}(S^{alg3}), f_{exp}(\{e^*\})\} \geq \frac{1-1/e}{2} f_{exp}(S^*)$.
\end{theorem}
\emph{Proof:} Let $R=\argmax_{S\in\{S^{alg3}, \{e^*\}\} }g_{exp}(S)$ denote the solution maximizing $g_{exp}(\cdot)$ between $S^{alg3}$ and  $\{e^*\}$.  Since $f_{exp}(R)\leq \max\{f_{exp}(S^{alg3}), f_{exp}(\{e^*\})\}$, to prove this theorem, it suffice to show that $f_{exp}(R) \geq  \frac{1-1/e}{2} f_{exp}(S^*)$.

First, Lemma \ref{lem:aaa1} implies that
\begin{eqnarray}
\label{eq:bbb1}
g_{exp}(R) \geq \frac{1-1/e}{2} g_{exp}(S^{p4}) \geq \frac{1-1/e}{2} g_{exp}(S^*)
 \end{eqnarray}
The second inequality is due to $S^*$ is a feasible solution to \textbf{P.4} and  $S^{p4}$ is the optimal solution to \textbf{P.4}.

 Second, due to $c(e^*) \leq c(S^*)$, which is because $e^*\in S^*$ and $c(S^{alg3})\leq c(S^*)$, which is because of the design of Algorithm \ref{alg:LPP12}, we have $B-c(e^*) \geq B-c(S^*)$ and $B-c(S^{alg1}) \geq B-c(S^*)$. Thus,
 \begin{eqnarray}
\label{eq:ccc1}B-c(R) \geq B-c(S^*)
 \end{eqnarray}
(\ref{eq:bbb1}) and (\ref{eq:ccc1}) imply that $f_{exp}(R)=\min\{g_{exp}(R), B-c(R)\} \geq \min\{\frac{1-1/e}{2}g_{exp}(S^*), B-c(S^*)\}$. Because $\frac{1-1/e}{2}\leq 1$, it follows that  $f_{exp}(R) \geq \frac{1-1/e}{2} \min\{g_{exp}(S^*), B-c(S^*)\} =\frac{1-1/e}{2}f_{exp}(S^*)$. $\Box$

\paragraph{Solving the General Seed Selection Problem}
\label{sec:unconstrained}
Now we are in position to study the general seed selection problem without knowing $c(S^*)$. Assume $|\mathcal{V}|=n$, i.e., there are $n$ users in the social network. The basic idea of our solution can be found as follows:
\begin{enumerate}
\item We first number all users in non-decreasing order of their costs. For each $i\in [n]$, let $\mathcal{V}_i=\{e_1, e_2, \cdots, e_i\}$ denote the first $i$ users.
\item For each $i\in [n]$, we build a set of subsets $\{S_{i,1}, S_{i,2}, \cdots, S_{i, i}\}$  iteratively according to the following benefit-cost greedy algorithm. It starts with $S_{i,0}=\emptyset$ and $k=0$. In each subsequent $k$, let
    \begin{eqnarray*}
S_{i,k+1}\leftarrow S_{i,k} \cup\{s_{i, k+1}\}
\end{eqnarray*}
where $s_{i, k+1}= \argmax_{e\in \mathcal{V}_i\setminus S_{i,k}} \frac{g_{exp}(S_{i,k}\cup\{e\})-g_{exp}(S_{i,k})}{c(e)}$ denotes the item in $\mathcal{V}_i\setminus S_{i,k}$ that maximizes the benefit-cost ratio with respect to $S_{i,k}$.
\item We also compute the expected utility for each single user and find the one $o$ maximizing the expected utility, i.e., $o=\argmax_{e\in \mathcal{V}} f_{exp}(\{e\})$.
\item Among all $\sum_{i=1}^n i + 1$ solutions returned from step 2 and step 3, we choose the one with the largest expected utility as the final solution.
\end{enumerate}

\begin{algorithm}[hptb]
\caption{Non-adaptive Seed Selection Algorithm for Deterministic Realization}
\label{alg:LPP13}
\begin{algorithmic}[1]
\STATE number all users in non-decreasing order of their costs
\FOR {$i=1$ to $n$}
\STATE let $\mathcal{V}_i=\{e_1, e_2, \cdots, e_i\}$ denote the top $i$ users
\STATE $S_{i,0}=\emptyset, k=0, U = \mathcal{V}_i$
\WHILE {$k\leq i$}
\STATE let $s_{i, k+1}$ denote the user $e$ maximizing $\frac{g_{exp}(S_{i,k}\cup\{e\})-g_{exp}(S_{i,k})}{c(e)}$ among all users in $U\setminus S_{i, k}$
\STATE let $S_{i, k+1}=S_{i, k}\cup\{s_{i, k+1}\}$
\STATE $k\leftarrow k+1$
\ENDWHILE
\ENDFOR
\STATE let $o\in \argmax_{e\in \mathcal{V}} f_{exp}(\{e\})$
\RETURN the best solution in $S_{1,1} \cup \{S_{2, 1}, S_{2, 2}\}\cup \cdots \cup \{S_{n,1} \cdots, S_{n,n}\}\cup\{o\}$
\end{algorithmic}
\end{algorithm}

A detailed description of our algorithm is listed in Algorithm \ref{alg:LPP13}.  We next analyze the approximation ratio of Algorithm \ref{alg:LPP13}. Let $S^{alg4}$ denote the solution returned from Algorithm \ref{alg:LPP13}.  Recall that $S^{alg3}$ is the solution returned from Algorithm \ref{alg:LPP12} which assumes the knowledge of $c(S^*)$. We first show that $f_{exp}(S^{alg4}) \geq f_{exp}(S^{alg3})$. Recall that $\mathcal{V}(c(S^*))=\{e\in \mathcal{V}\mid c(e)\leq c(S^*)\}$ represents the set of users whose cost is no larger than $c(S^*)$.
Assume $|\mathcal{V}(c(S^*))|=m$, then we have $\mathcal{V}_m=\mathcal{V}(c(S^*))$, i.e., $\mathcal{V}(c(S^*))$ is composed of the top $m$ users with the least cost. Moreover, assume $|S^{alg3}|= r$, i.e., Algorithm \ref{alg:LPP12} selects $r$ users.  Recall that $S_{m,r}$ is the first $r$ users selected by the benefit-cost greedy algorithm subject to a ground set $\mathcal{V}_m$. Because $\mathcal{V}_m=\mathcal{V}(c(S^*))$ and  $|S^{alg3}|= r$, the output returned from Algorithm \ref{alg:LPP12} is identical to $S_{m,r}$, i.e.,  $S_{m,r}=S^{alg3}$. It follows that
    \begin{eqnarray}
    \label{eq:ddd1}
f_{exp}(S^{alg4}) \geq f_{exp}(S_{m,r})=f_{exp}(S^{alg3})
\end{eqnarray}
Moreover, because $f_{exp}(S^{alg4})\geq f_{exp}(\{o\})$ and $f_{exp}(\{o\})\geq f_{exp}(\{e^*\})$, we have $f_{exp}(S^{alg4}) \geq f_{exp}(\{e^*\})$. This together with (\ref{eq:ddd1}) implies that
    \begin{eqnarray}
    \label{eq:dddd1}
f_{exp}(S^{alg4}) \geq \max\{f_{exp}(S^{alg3}), f_{exp}(\{e^*\})\}
\end{eqnarray}
The following theorem follows from Theorem \ref{thm:11} and (\ref{eq:dddd1}).
\begin{theorem}
Let $S^{alg4}$ denote the solution returned from Algorithm \ref{alg:LPP13}, $f_{exp}(S^{alg4}) \geq \frac{1-1/e}{2} f_{exp}(S^*)$.
\end{theorem}

\section{Adaptive Seed Selection Problem}

\begin{algorithm}[hptb]
\caption{Greedy Policy $\pi^1$}
\label{alg:LPP4}
\begin{algorithmic}[1]
\STATE $t=0; S_0=\emptyset; \psi_0=\emptyset; U=\mathcal{V}$.
\WHILE {$U\setminus S_{t}\neq \emptyset$}
\STATE let $s_{t+1} = \argmax_{e\in U\setminus S_{t}} \frac{\Delta_{h(\cdot, \cdot, 0)}(e\mid \psi_{t})}{c(e)}$;
\IF {$c(s_{t+1}) + c(S_{t}) \leq C$}
\STATE $S_{t+1}\leftarrow S_{t}\cup\{e_{t+1}\}$
\STATE $\psi_{t+1}\leftarrow \psi_{t}\cup\{(s_{t+1}, \Phi(s_{t+1}))\}$;
\STATE $t\leftarrow t+1$;
\ELSE
\RETURN $S_{t}$
\ENDIF
\ENDWHILE
\RETURN $S_t$
\end{algorithmic}
\end{algorithm}

In this section, we study the seed selection problem under the adaptive setting. Before presenting our adaptive policy, we first introduce some additional notations. Given any number $z\in [0, B]$, define $h(\cdot, \cdot, z): 2^{\mathcal{V}}\times O^\mathcal{V}\rightarrow \mathbb{R}_{\geq0}$ as follows:
\begin{eqnarray*}
h(S, \phi, z) = \min\{g(S, \phi), B-z\}
\end{eqnarray*}
And the expected utility of a policy $\pi$ under $h(\cdot, \cdot, z)$ is defined as
\begin{eqnarray*}
h_{avg}(\pi, z) = \mathbb{E}[h(\mathcal{V}(\pi, \Phi), \Phi, z)] = \mathbb{E}[\min\{g(\mathcal{V}(\pi, \Phi), \Phi), B-z\}]
\end{eqnarray*}

For any $z\in [0, B]$ and partial realization $\psi$, let $\Delta_{h(\cdot, \cdot, z)}(e\mid \psi)$ denote the marginal utility of $e$ on top of $\psi$ under $h(\cdot, \cdot, z)$, i.e.,
\begin{eqnarray*}
\Delta_{h(\cdot, \cdot, z)}(e\mid \psi) = \mathbb{E}_{\Phi\sim \psi}[h(\mathrm{dom}(\psi)\cup\{e\}, \Phi, z)]-\mathbb{E}_{\Phi\sim \psi}[h(\mathrm{dom}(\psi), \Phi, z)]
\end{eqnarray*}

Let $\overline{e}$ denote the most expensive user in $\mathcal{V}$, i.e., $\overline{e}\in \argmax_{e\in \mathcal{V}} c(e)$. Let $C=\max\{c(\overline{e}), \frac{B}{2}\}$.  Now we are ready to introduce our \emph{Adaptive Seed Selection Policy} $\pi^s$.
\paragraph{Design of Adaptive Seed Selection Policy $\pi^s$.}  The design of $\pi^s$ involves two candidate policies:  $\pi^1$ and $\pi^2$. And $\pi^s$ samples a policy uniformly at random from $\{\pi^1, \pi^2\}$.  Thus, the expected utility $f_{avg}(\pi^s)$ of $\pi^s$ can be represented as $f_{avg}(\pi^s)=\frac{f_{avg}(\pi^1)+f_{avg}(\pi^2)}{2}$. We next describe the design of $\pi^1$ and $\pi^2$ in details.
\begin{itemize}
\item \textbf{Design of $\pi^1$.} The first candidate $\pi^1$ (Algorithm \ref{alg:LPP4}) is an adaptive version of benefit-cost greedy algorithm presented in the earlier section.  $\pi^1$ starts with $t=0$, an initial solution $S_0=\emptyset$ and initial partial realization $\psi_0=\emptyset$. In each iteration $t+1$, it adds $s_{t+1}$ to the current solution $S_t$. i.e., $S_{t+1}\leftarrow  S_{t} \cup\{s_{t+1}\}$,
where
\[s_{t+1}= \argmax_{e\in U\setminus S_{t}} \frac{\Delta_{h(\cdot, \cdot, 0)}(e\mid \psi_{t})}{c(e)}\]  is the user maximizing the benefit-cost ratio with respect to the current observation $\psi_t$ under $h(\cdot, \cdot, 0)$. Then we update the observation using $\psi_{t+1}\leftarrow \psi_{t}\cup\{(s_{t+1}, \Phi(s_{t+1}))\}$ and enter the next iteration. This process iterates until it reaches some iteration $t$ such that $c(S_{t+1})+c(s_{t+1})> C$. Recall that $C=\max\{c(\overline{e}), \frac{B}{2}\}$.
\item \textbf{Design of $\pi^2$.} The second candidate $\pi^2$ simply returns the singleton $e$ that maximizes $f_{avg}(\{e\})$.
\end{itemize}

\paragraph{Performance Analysis}
We next analyze the approximation ratio of $\pi^s$. All missing proofs are moved to appendix. We start with two technical lemmas. We first show that $f_{avg}(\pi) \geq h_{avg}(\pi, 0)$ for any $\pi$.
\begin{lemma}
\label{lem:monday}
For any policy $\pi$, we have $f_{avg}(\pi) \geq h_{avg}(\pi, 0)$.
\end{lemma}

We next prove that $h(\cdot, \cdot, z): 2^\mathcal{V}\times O^\mathcal{V}\rightarrow \mathbb{R}_{\geq0}$  is adaptive monotone and adaptive submodular for any $z\in [0, B]$.
\begin{lemma}
\label{lem:adaptivesubmodular}
For any $z\in [0, B]$, $h(\cdot, \cdot, z): 2^\mathcal{V}\times O^\mathcal{V}\rightarrow \mathbb{R}_{\geq0}$  is adaptive monotone and adaptive submodular.
\end{lemma}

Now we are ready to present the main theorem.
\begin{theorem}
\label{lem:sunny}
Let $C=\max\{c(\overline{e}), \frac{B}{2}\}$ and $\alpha=\min\{\frac{1}{2}, 1-\frac{C}{B}\}$, and let $\pi^*$ denote the optimal policy for the adaptive seed selection problem, we have  $f_{avg}(\pi^s) \geq \alpha \frac{1-e^{-\frac{C}{B}}}{2} f_{avg}(\pi^*)$.
\end{theorem}
Notice when $c(\overline{e}) \leq \frac{B}{2}$, i.e., the cost of the most expensive user is no larger than $B/2$, the above approximation ratio is lower bounded by $\frac{1-e^{-\frac{1}{2}}}{4}$.

\section{Conclusion}
In this paper, we study both non-adaptive and adaptive seed selection problem in the context of incentivized social advertising. We develop effective solutions for both cases. Although the focus of this paper is on  seed selection problem, the theoretical contributions made in this work apply to a broad range of optimization problems whose objective function might take on negative values. In the future, we would like to extend this study by considering multiple advertisers.
\bibliographystyle{ijocv081}
\bibliography{reference}
\section{Appendix}
\subsection{Proof of Lemma \ref{lem:monday}}
\emph{Proof:}  Recall that $f_{avg}(\pi)=\mathbb{E}[f(\mathcal{V}(\pi, \Phi), \Phi)]=\mathbb{E}[\min\{g(\mathcal{V}(\pi, \Phi), \Phi), B-c(\mathcal{V}(\pi, \Phi))\}]$. We also have $h_{avg}(\pi, 0)=\mathbb{E}[h(\mathcal{V}(\pi, \Phi), \Phi, 0)]=\mathbb{E}[\min\{g(\mathcal{V}(\pi, \Phi), \Phi), B\}]$. Because $\mathbb{E}[\min\{g(\mathcal{V}(\pi, \Phi), \Phi), B-c(\mathcal{V}(\pi, \Phi))\}]\leq \mathbb{E}[\min\{g(\mathcal{V}(\pi, \Phi), \Phi), B\}]$, we have $f_{avg}(\pi) \geq h_{avg}(\pi, 0)$. $\Box$

\subsection{Proof of Lemma \ref{lem:adaptivesubmodular}}
\emph{Proof:} The proof of the adaptive monotonicity of  $h(\cdot, \cdot, z): 2^\mathcal{V}\times O^\mathcal{V}\rightarrow \mathbb{R}_{\geq0}$  is trivial. 
Because $g(\cdot, \phi)$ is monotone for any realization $\phi$, we have $g(\{e\}\cup\mathrm{dom}(\psi), \phi)\geq g(\mathrm{dom}(\psi), \phi)$. Thus, $\mathbb{E}_{\Phi\sim \psi}[\min\{g(\{e\}\cup\mathrm{dom}(\psi), \Phi), B-z\}-\min\{g(\mathrm{dom}(\psi), \Phi), B-z\}] \geq 0$, it follows that $\Delta_{h(\cdot, \cdot, z)} (e\mid \psi)\geq 0$.

We next focus on proving the adaptive submodularity of  $h: 2^\mathcal{V}\times O^\mathcal{V}\rightarrow \mathbb{R}_{\geq0}$. Consider any two partial realizations $\psi^b$ and $\psi^a$ such that $\psi^b \subseteq \psi^a$ and  a user $e\notin \mathrm{dom}(\psi^a)$. Let $\mathcal{V}'=\mathcal{V}\setminus \mathcal{V}(\psi^a)$ and $\mathcal{E}'=\mathcal{E} \setminus\mathcal{E}(\psi^a)$ where $\mathcal{V}(\psi^a)$ is the set of engaged users under $\psi^a$ and $\mathcal{E}(\psi^a)$ is the set of edges whose label (\verb"Live" or \verb"Blocked") can be observed under $\psi^a$. For each $v\in \mathcal{V}'$ and a realization $\phi$, we define the $\mathcal{E}'$-restricted realization $\theta(\phi)(u)$ of $u$ as a function $\theta(\phi)(u): \mathcal{E}' \rightarrow \{\verb"Blocked", \verb"Live", ?\}$ such that for each edge $(v, w)\in \mathcal{E}'$, if $\phi(u)((v,w))=\verb"Live"$ and  $v$  can be reached through some path composed of live edges from $\mathcal{E}'$, then set $\theta(\phi)(u)((v,w))=\verb"Live"$; if $\phi(u)((v,w))=\verb"Blocked"$ and  $v$ can be reached through some path composed of live edges from $\mathcal{E}'$, then set $\theta(\phi)(u)((v,w))=\verb"Blocked"$; otherwise, set $\theta(\phi)(u)((v,w))=?$. Intuitively,  given a realization $\phi$, $\theta(\phi)(u)$ contains the labels of those edges whose labels can be observed after $u$ is being selected even if all edges in $\mathcal{E}(\psi^a)$ are removed from $G$. Let $\theta(\Phi)(u)$ denote a random realization of the $\mathcal{E}'$-restricted realization of $u$ and let $\theta(\Phi)=\{\theta(\Phi)(u)\mid u\in \mathcal{V}'\}$. Consider a fixed $\theta(\phi)$, a full realization $\phi'$, and any user $u\in \mathcal{V}'$, we say $\theta(\phi)(u)$ is consistent with $\phi'(u)$ if they are equal everywhere in the domain of edges whose labels can be observed under both $\theta(\phi)$ and $\phi'$.

 Now consider a fixed $\theta(\phi)$ and any $e \in \mathcal{V}'$, let $\phi^a$ be the \emph{fixed} full realization that is consistent with $\psi^a\cup \theta(\phi)$, i.e., $\phi^a$ is fixed once $\psi^a$ and $\theta(\phi)$ are given, and let $\phi^b$ be \emph{any} realization that is consistent with $\psi^b\cup \theta(\phi)$, we have
\begin{eqnarray}
\label{eq:fff}
g(\mathrm{dom}(\psi^a)\cup\{e\}, \phi^a)-g(\mathrm{dom}(\psi^a), \phi^a) \leq g(\mathrm{dom}(\psi^b)\cup\{e\}, \phi^b)-g(\mathrm{dom}(\psi^b), \phi^b)
\end{eqnarray}
This is because for any user $u\in \mathcal{V}'$, if $u$ is influenced by $e$ under $\phi^a$, which implies that there exists a path composed of live edges in $\mathcal{E}'$ such that it connects $u$ with $e$, then $u$ must be influenced  by $e$ conditional on $\phi^b$ due to such a path can also be found in $\phi^b$ since $\phi^b$ is consistent with $\theta(\phi)$. Moreover, we have $g(\mathrm{dom}(\psi^a), \phi^a) \geq g(\mathrm{dom}(\psi^b), \phi^b)$ due to $\psi^b \subseteq \psi^a$. This together with (\ref{eq:fff}) and Lemma \ref{lem:end} (in appendix) implies that for any $z\in [0, B]$, $\min\{g(\mathrm{dom}(\psi^a)\cup\{e\}, \phi^a), B-z\}-\min\{g(\mathrm{dom}(\psi^a), \phi^a), B-z\}\leq \min\{g(\mathrm{dom}(\psi^b)\cup\{e\}, \phi^b), B-z\}-\min\{g(\mathrm{dom}(\psi^b), \phi^b), B-z\}$. Thus, for any $\phi^a$, $\phi^b$ and $\theta(\phi)$ such that $\phi^a\sim \psi^a\cup \theta(\phi)$ and $\phi^b\sim \psi^b\cup \theta(\phi)$ and any $z\in [0, B]$, we have
\begin{eqnarray}
&&h(\mathrm{dom}(\psi^a)\cup\{e\}, \phi^a, z)-h(\mathrm{dom}(\psi^a), \phi^a, z) ~\nonumber\\
&&= \min\{g(\mathrm{dom}(\psi^a)\cup\{e\}, \phi^a), B-z\}-\min\{g(\mathrm{dom}(\psi^a), \phi^a), B-z\}~\nonumber\\
&&\leq \min\{g(\mathrm{dom}(\psi^b)\cup\{e\}, \phi^b), B-z\}-\min\{g(\mathrm{dom}(\psi^b), \phi^b), B-z\}~\nonumber\\
&&=h(\mathrm{dom}(\psi^b)\cup\{e\}, \phi^b, z)-h(\mathrm{dom}(\psi^b), \phi^b, z) \label{eq:endend}
\end{eqnarray}

It follows that for any $\theta(\phi)$ and  $z\in [0, B]$,
\begin{eqnarray}
&&\mathbb{E}_{\Phi\sim \psi^a\cup \theta(\phi)}[h(\mathrm{dom}(\psi^a)\cup\{e\}, \Phi, z)-h(\mathrm{dom}(\psi^a), \Phi, z)] ~\nonumber\\
&&= \sum_{\phi'}\Pr[\Phi=\phi'\mid \Phi\sim \psi^a\cup \theta(\phi) ] (h(\mathrm{dom}(\psi^a)\cup\{e\}, \phi', z)-h(\mathrm{dom}(\psi^a), \phi', z)) ~\nonumber\\
&&\leq \sum_{\phi'}\Pr[\Phi=\phi'\mid \Phi\sim \psi^b\cup \theta(\phi) ] (h(\mathrm{dom}(\psi^b)\cup\{e\}, \phi', z)-h(\mathrm{dom}(\psi^b), \phi', z)) ~\nonumber\\
&&= \mathbb{E}_{\Phi\sim \psi^b\cup \theta(\phi)}[h(\mathrm{dom}(\psi^b)\cup\{e\}, \Phi, z)-h(\mathrm{dom}(\psi^b), \Phi, z)]\label{eq:000}
\end{eqnarray}
The inequality is due to $\sum_{\phi'}\Pr[\Phi=\phi'\mid \Phi\sim \psi^a\cup \theta(\phi) ] =1$ and $\sum_{\phi'}\Pr[\Phi=\phi'\mid \Phi\sim \psi^b\cup \theta(\phi) ] =1$, and (\ref{eq:endend}) holds  for any $\phi^a$, $\phi^b$ and $\theta(\phi)$ such that $\phi^a\sim \psi^a\cup \theta(\phi)$ and $\phi^b\sim \psi^b\cup \theta(\phi)$ and any $z\in [0, B]$.


Moreover, because edges are labeled independently, $\theta(\Phi)$ is drawn from the same distribution under both $\psi^b$ and $\psi^a$. Thus, we can represent $\Delta_{h(\cdot, \cdot, z)} (e\mid \psi^a)$ and $\Delta_{h(\cdot, \cdot, z)} (e\mid \psi^b)$ as follows.
\begin{eqnarray*}
&&\Delta_{h(\cdot, \cdot, z)} (e\mid \psi^a)=\mathbb{E}_{\theta(\Phi)}[\mathbb{E}_{\Phi'\sim \psi^a\cup \theta(\Phi)}[h(\mathrm{dom}(\psi^a)\cup\{e\}, \Phi', z)]]
\end{eqnarray*}
\begin{eqnarray*}
&&\Delta_{h(\cdot, \cdot, z)} (e\mid \psi^b)=\mathbb{E}_{\theta(\Phi)}[\mathbb{E}_{\Phi'\sim \psi^b\cup \theta(\Phi)}[h(\mathrm{dom}(\psi^b)\cup\{e\}, \Phi',  z)]]
\end{eqnarray*}
This together with (\ref{eq:000}) implies that $\Delta_{h(\cdot, \cdot, z)} (e\mid \psi^a) \leq \Delta_{h(\cdot, \cdot, z)} (e\mid \psi^b)$. $\Box$

\subsection{Proof of Theorem \ref{lem:sunny}}
\emph{Proof:}
We first introduce a new optimization problem \textbf{P.5}.
\begin{center}
\framebox[0.78\textwidth][c]{
\enspace
\begin{minipage}[t]{0.45\textwidth}
\small
\textbf{P.5:} \emph{Maximize $h_{avg}(\pi, 0)$}\\
\textbf{subject to:}
\begin{equation*}
\begin{cases}
\forall \phi: \mathcal{V}(\pi, \phi)\subseteq \mathcal{V}\\
\forall \phi: c(\mathcal{V}(\pi, \phi))\leq  B
\end{cases}
\end{equation*}
\end{minipage}
}
\end{center}
Let $\pi^{p5}$ denote the optimal policy for \textbf{P.5}. Clearly, the optimal policy $\pi^*$ is a feasible solution for \textbf{P.5}, i.e., it satisfies both constraints listed in \textbf{P.5}. Thus, $h_{avg}(\pi^*, 0) \leq h_{avg}(\pi^{p5}, 0)$. Moreover, due to $h_{avg}(\pi^*, 0)\geq f_{avg}(\pi^*)$ (Lemma \ref{lem:monday}), we have $ h_{avg}(\pi^{p5}, 0) \geq f_{avg}(\pi^*)$. Thus, to prove this theorem, it suffice to show that $f_{avg}(\pi^s) \geq \alpha \frac{1-e^{-\frac{C}{B}}}{2} h_{avg}(\pi^{p5}, 0)$.

Recall that $\pi^1$ (Algorithm \ref{alg:LPP4}) is an adaptive benefit-cost greedy algorithm. In each iteration, it adds a user maximizing the benefit-cost ratio with respect to the current observation to the solution, and it terminates right before violating the budget constraint $C$. Now consider a ``one-step-further'' version $\pi^{1+}$ of $\pi^1$ which is obtained by first running $\pi^1$ then selecting one more user according to the same greedy manner. One can verify that $\pi^{1+}$ always violates the budget $C$ (assuming $c(\mathcal{V})> C$ to avoid trivial cases, otherwise, one can simply select all users from $\mathcal{V}$ to obtain an optimal solution).  Theorem 2 in \cite{tang2020influence} states that if $h(\cdot, \cdot, 0): 2^\mathcal{V}\times O^\mathcal{V}\rightarrow \mathbb{R}_{\geq0}$ is adaptive monotone and adaptive submodular,  which is proved in Lemma \ref{lem:adaptivesubmodular}, then running $\pi^{1+}$ which consumes at least $\frac{C}{B}$ fraction of the budget $B$ achieves a $1-e^{-\frac{C}{B}}$ approximation ratio, i.e.,
\begin{eqnarray}
\label{eq:lastlast1}
h_{avg}(\pi^{1+}, 0) \geq  (1-e^{-\frac{C}{B}})h_{avg}(\pi^{p5}, 0)
 \end{eqnarray}
Moreover, due to $h(\cdot, \cdot, 0): 2^\mathcal{V}\times O^\mathcal{V}\rightarrow \mathbb{R}_{\geq0}$ is adaptive submodular, we have for any partial realization $\psi$ and user $e\in \mathcal{V}$, we have $\Delta_{h(\cdot, \cdot, 0)}(e \mid \psi)\leq \Delta_{h(\cdot, \cdot, 0)}(\{e\} \mid \emptyset)\leq \max_{e\in \mathcal{V}}\Delta_{h(\cdot, \cdot, 0)}(\{e\} \mid \emptyset)\leq \max_{e\in \mathcal{V}} h_{avg}(\{e\}, 0)$. Thus, the marginal utility brought by the last user of $\pi^1$ under  $h(\cdot, \cdot, 0): 2^\mathcal{V}\times O^\mathcal{V}\rightarrow \mathbb{R}_{\geq0}$ is at most $\max_{e\in \mathcal{V}} h_{avg}(\{e\}, 0)$. It follows that $h_{avg}(\pi^{1}, 0)+\max_{e\in \mathcal{V}} h_{avg}(\{e\}, 0) \geq h_{avg}(\pi^{1+}, 0)$. This together with (\ref{eq:lastlast1}) implies
 \begin{eqnarray}
\label{eq:lastlast}
h_{avg}(\pi^{1}, 0)+\max_{e\in \mathcal{V}} h_{avg}(\{e\}, 0)  \geq  (1-e^{-\frac{C}{B}})h_{avg}(\pi^{p5}, 0)
 \end{eqnarray}

 Let $\alpha=\min\{\frac{1}{2}, 1-\frac{C}{B}\}$. We next show that  $ \alpha h_{avg}(\pi^1, 0) \leq f_{avg}(\pi^1)$ and  $\alpha \cdot \max_{e\in \mathcal{V}} h_{avg}(\{e\}, 0) \leq  f_{avg}(\pi^2)$. First,
\begin{eqnarray}
f_{avg}(\pi^1)&&=\mathbb{E}[f(\mathcal{V}(\pi^1, \Phi), \Phi)]~\nonumber\\
&&=\mathbb{E}[\min\{g(\mathcal{V}(\pi^1, \Phi), \Phi), B-c(\mathcal{V}(\pi^1, \Phi))\}]~\nonumber\\
&&\geq \mathbb{E}[\min\{g(\mathcal{V}(\pi^1, \Phi), \Phi), B-C\}]~\nonumber\\
&&\geq \mathbb{E}[\min\{g(\mathcal{V}(\pi^1, \Phi), \Phi), \alpha B\}]~\nonumber\\
&&\geq \alpha \mathbb{E}_[\min\{g(\mathcal{V}(\pi^1, \Phi), \Phi), B\}] = \alpha h_{avg}(\pi^1, 0) \label{eq:99}
\end{eqnarray}
The first inequality is due to  $\forall \phi: c(\mathcal{V}(\pi^1, \phi))\leq C$, i.e., $\pi^1$ never violates the budget constraint $C$.  The second inequality is due to $\alpha=\min\{\frac{1}{2}, 1-\frac{C}{B}\}$. The third inequality is due to $\alpha \leq 1$. Now we follow the same procedure to prove that  $\alpha \cdot \max_{e\in \mathcal{V}} h_{avg}(\{e\}, 0) \leq  f_{avg}(\pi^2)$.  Let $u \in \argmax_{e\in \mathcal{V}} f_{avg}(\{e\})$ and $w \in \argmax_{e\in \mathcal{V}} h_{avg}(\{e\}, 0)$.  Because $\pi^2$ selects the singleton $u$, we have $f_{avg}(\pi^2)= f_{avg}(\{u\})$.
It follows that
\begin{eqnarray}
f_{avg}(\pi^2)&&= f_{avg}(\{u\}) \geq f_{avg}(\{w\})~\nonumber\\
&&= \mathbb{E}[\min\{g(\{w\}, \Phi), B-c(w)\}]~\nonumber\\
&&\geq \mathbb{E}[\min\{g(\{w\}, \Phi), B-C\}]~\nonumber\\
&&\geq \mathbb{E}[\min\{g(\{w\}, \Phi), \alpha B\}] ~\nonumber\\
&&\geq \alpha \mathbb{E}[\min\{g(\{w\}, \Phi), B\}] ~\nonumber\\
&&= \alpha h_{avg}(\{w\}, 0)= \alpha\cdot \max_{e\in \mathcal{V}} h_{avg}(\{e\}, 0)\label{eq:opop}
\end{eqnarray}
The first inequality is due to the definition of $u$. The second inequality is due to $c(w) \leq C$.   The third inequality is due to $\alpha=\min\{\frac{1}{2}, 1-\frac{C}{B}\}$. The forth inequality is due to $\alpha \leq 1$. The last equality is due to the definition of $w$. 
(\ref{eq:99}) and (\ref{eq:opop}) imply that  $ f_{avg}(\pi^1)+ f_{avg}(\pi^2) \geq  \alpha (h_{avg}(\pi^1, 0) +  \max_{e\in \mathcal{V}} h_{avg}(\{e\}, 0)) $. This together with (\ref{eq:lastlast}) implies that
\begin{eqnarray}
\label{eq:9911}
f_{avg}(\pi^1)+ f_{avg}(\pi^2) \geq \alpha(1-e^{-\frac{C}{B}})h_{avg}(\pi^{p5}, 0)
 \end{eqnarray}

Recall that $\pi^s$  samples a policy uniformly at random from $\{\pi^1, \pi^2\}$.  Thus, the expected utility $f_{avg}(\pi^s)$ of $\pi^s$ is $f_{avg}(\pi^s)=\frac{f_{avg}(\pi^1)+f_{avg}(\pi^2)}{2}$. This together with (\ref{eq:9911}) implies that  $f_{avg}(\pi^s) \geq \frac{\alpha (1-e^{-\frac{C}{B}})}{2} h_{avg}(\pi^{p5}, 0)$. $\Box$
\subsection{Lemma \ref{lem:end}}
\begin{lemma}
\label{lem:end}
Consider any five constants $c_1, c_2, c_3, c_4$ and $x$ such that $c_1\geq c_2$ and $c_3 \geq c_4$, $c_1-c_2\geq c_3-c_4$ and $c_2 \leq c_4$, we have $\min\{c_1, x\}-\min\{c_2, x\}\geq\min\{c_3, x\}-\min\{c_4, x\}$.
\end{lemma}
\emph{Proof:}  For the case when $c_1 \leq c_3$, this result has been proved in Lemma 2 in \cite{tang2016optimizing}. We next focus on the case when $c_1 > c_3$. We prove this lemma in five subcases depending on relation between $x$ and the other four constants. Notice that when $c_1 > c_3$, $c_1\geq c_2$, $c_3 \geq c_4$, and $c_2 \leq c_4$, we have $c_1 > c_3 \geq c_4 \geq c_2$.
\begin{itemize}
\item If $x \geq c_1 > c_3 \geq c_4 \geq c_2$, then $\min\{c_1, x\}=c_1$, $\min\{c_2, x\}=c_2$, $\min\{c_3, x\}=c_3$ and $\min\{c_4, x\}=c_4$. Thus, $\min\{c_1, x\}-\min\{c_2, x\}\geq\min\{c_3, x\}-\min\{c_4, x\}$ due to the assumption that $c_1-c_2\geq c_3-c_4$.
\item If $c_1 > x \geq c_3 \geq c_4 \geq c_2$, then $\min\{c_1, x\}=x$, $\min\{c_2, x\}=c_2$, $\min\{c_3, x\}=c_3$ and $\min\{c_4, x\}=c_4$. Thus, $\min\{c_1, x\}-\min\{c_2, x\} = x- c_2$ and $\min\{c_3, x\}-\min\{c_4, x\}=c_3-c_4$. Because $x\geq c_3$ and $c_2\leq c_4$, we have $x- c_2 \geq c_3-c_4$. It follows that $\min\{c_1, x\}-\min\{c_2, x\}\geq\min\{c_3, x\}-\min\{c_4, x\}$.
\item If $c_1 >c_3 > x \geq  c_4 \geq c_2$, then $\min\{c_1, x\}=x$, $\min\{c_2, x\}=c_2$, $\min\{c_3, x\}=x$ and $\min\{c_4, x\}=c_4$. Thus, $\min\{c_1, x\}-\min\{c_2, x\} = x- c_2$ and $\min\{c_3, x\}-\min\{c_4, x\}=x-c_4$. Because $c_2 \leq c_4$, we have $x- c_2 \geq x-c_4$, thus, $\min\{c_1, x\}-\min\{c_2, x\}\geq\min\{c_3, x\}-\min\{c_4, x\}$.
\item If $c_1 >c_3 \geq  c_4 > x  \geq c_2$, then $\min\{c_1, x\}=x$, $\min\{c_2, x\}=c_2$, $\min\{c_3, x\}=x$ and $\min\{c_4, x\}=x$. Thus, $\min\{c_1, x\}-\min\{c_2, x\} = x- c_2$ and $\min\{c_3, x\}-\min\{c_4, x\}=x-x =0$. Because $x  \geq c_2$, we have $x- c_2 \geq 0$, thus, $\min\{c_1, x\}-\min\{c_2, x\}\geq\min\{c_3, x\}-\min\{c_4, x\}$.
\item If $c_1 >c_3 \geq  c_4 \geq c_2 > x  $, then $\min\{c_1, x\}=x$, $\min\{c_2, x\}=x$, $\min\{c_3, x\}=x$ and $\min\{c_4, x\}=x$. Thus, $\min\{c_1, x\}-\min\{c_2, x\} = x- x = 0$ and $\min\{c_3, x\}-\min\{c_4, x\}=x-x =0$. Thus, $\min\{c_1, x\}-\min\{c_2, x\}\geq\min\{c_3, x\}-\min\{c_4, x\}$.
\end{itemize}

$\Box$

\end{document}